\documentclass[%
 reprint,
nofootinbib,
 amsmath,amssymb,
 aps,
]{revtex4-1}

\usepackage{graphicx}
\usepackage{dcolumn}
\usepackage{bm}
\usepackage{xifthen}
\usepackage{times}
\usepackage{xcolor}

\newcommand{\Eqref}[1]{Eq.\ \eqref{#1}}
\newcommand{\mm}{m}
\newcommand{\MM}{\mm (\vec{x} , t)}
\renewcommand*{\vec}[1]{\boldsymbol{#1}}
\newcommand{\dd}{{\rm d}}
\newcommand{\EE}{E}
\newcommand{\px}{\rho}
\newcommand{\pxp}{p}
\newcommand{\PXP}{\pxp (\vec{x}, \phi | t)}
\newcommand{\PXPper}{\bar\pxp (\vec{x}, \phi | t)}
\newcommand{\phxp}{\hat{\pxp}}
\newcommand{\PHXP}{\phxp (\vec{x}, \lambda | t)}
\newcommand{\PP}{P}
\newcommand{\ssdx}{\px_{\rm st}}
\newcommand{\ssdxp}{\pxp_{\rm st}}
\newcommand{\OMG}{\omega ( \vec{x}, \phi, t )}
\newcommand{\OMGx}{\omega (\vec{x}, t)}
\newcommand{\del}{\partial}
\newcommand{\DD}{\mathbf{D}}
\newcommand{\Difop}{\Delta^{\!\rm Br}}
\newcommand{\UU}{V}
\newcommand{\kB}{k_{\rm B}}
\newcommand{\BZ}{\Bz (\vec{x},t)}
\newcommand{\Bz}{b}
\newcommand{\flexmath}[4]{
  \ifthenelse{\equal{#1}{inline}}
             {$#2$#4}
             {\begin{#1}#2 \,#4\label{#3}\end{#1}}%
}
\newcommand{\gw}{\vec{g}}
\newcommand{\qq}{\vec{q}}
\newcommand{\mT}{{\sf T}}
\newcommand{\Tr}{\mathop{\mathrm{Tr}}}
\newcommand{\CC}{\mathbf{C}}
\newcommand{\QQ}{\vec{Q}}

\newcommand{\BB}{\mathbf{B}}
\newcommand{\PPPDlong}{position-phase probability density}
\newcommand{\PPDxlong}{position-phase (probability) density}
\newcommand{\PPD}{PPD}

\begin{document}


\title{Dynamics of \PPPDlong\ in magnetic resonance}

\author{Cem Yolcu}
\affiliation{Dept.\ of Biomedical Engineering, Link\"oping University, Sweden}
\author{Magnus Herberthson}%
\affiliation{Dept.\ of Mathematics, Link\"oping University, Sweden}
\author{Carl-Fredrik Westin}
\affiliation{Dept.\ of Radiology, Brigham and Women's Hospital, Harvard Medical School, Boston, MA}
\author{Evren \"Ozarslan}
\affiliation{Dept.\ of Biomedical Engineering, Link\"oping University, Sweden}
\affiliation{Center for Medical Image Science and Visualization, Link\"oping University, Sweden.}


\date{\today}

\begin{abstract}
We consider the behaviour of precessional angle (phase) carried by
molecules of a diffusing specimen under magnetic fields typical of
magnetic resonance experiments. An evolution equation for the ensemble
of particles is constructed, which treats the phase as well as the
position of the molecules as random variables. This ``\PPDxlong''
(\PPD) is shown to encode solutions to a family of Bloch-Torrey
equations (BTE) for transverse magnetization density, which is because
the \PPD\ is a more fundamental quantity than magnetization density;
the latter emerges from the former upon averaging. The present
paradigm represents a conceptual advantage, since the \PPD\ is a true
probability density subject to Markovian dynamics, rather than an
aggregate magnetization density whose evolution is less intuitive. We
also work out the analytical solution for suitable special cases.
\end{abstract}

\pacs{Valid PACS appear here}
\maketitle


Nuclear magnetic resonance (NMR) experiments detect radiation
originating from the Larmor precession of nuclear magnetic moments
around a polarizing magnetic field, $\vec{B}_0$
\cite{Abragambook}. Therefore the density $\MM$ of magnetic moments in
a piece of material has traditionally played the role of the
fundamental quantity whence the observed signal emerges. When $\MM$ is
treated as a complex number representing the components of
magnetization \emph{transverse} to $\vec{B}_0$ in a coordinate frame
rotating at the Larmor precession rate $\omega_0 = |\gamma
{B}_0|$,\footnote{$\gamma$ denotes the gyromagnetic ratio of the
  nucleus; the ratio of its magnetic moment to its spin.} the signal
amplitude arises as the integrated magnetization, \flexmath{align}
{\EE(t) = \textstyle\int \dd^3x \, \MM} {eq:signal}{,} over the region of
interest.

Under a spatially inhomogeneous magnetic field, nuclei experience
different precession rates at different locations. Therefore,
molecules of a fluid following a statistical distribution of paths
accumulate a distribution of precession phases, resulting in a reduced
transverse magnetization with respect to a coherent ensemble of
precessors. Such reduction of magnetization, hence the reduction in
signal, is widely used and investigated to quantify diffusive motion
in materials and biological tissues, as it contains signatures of the
structure of the microscopic environment which the fluid
inhabits \cite{PriceBook,Callaghanbook2}. Torrey
\cite{Torrey56} extended the differential equation of Bloch
\cite{Bloch46} that describes the rotation of magnetic moments (spins)
in an applied magnetic field to account for the diffusive motion of
the spin-carrying molecules (such as water), culminating in the
Bloch-Torrey equation (BTE) that determines the time evolution of the
magnetization density $\MM$.

With the phenomenological relaxation factor ${e}^{-t/T_2}$ divided out
of $\MM$, and $\BZ$ denoting the spatially inhomogeneous part of the
longitudinal magnetic field, the BTE reads
\begin{align}
  \del_t \MM = \Difop \MM - {i} \gamma \BZ \MM \,. \label{eq:BTE} 
\end{align}
Here the diffusion operator is given by
\begin{align}
  \Difop = \nabla \cdot e^{- \UU(\vec{x})} \DD (\vec{x}) \cdot
  \nabla {e}^{ \UU(\vec{x})} \,, \label{eq:Sm.op}
\end{align}
where $\UU(\vec{x})$ is the potential energy field normalized by the
thermal energy $\kB T$, and $\DD$ is a generally-anisotropic
diffusivity tensor.

We note that $\MM$ is an \emph{average} quantity. The molecules
that arrive in the vicinity of $\vec{x}$ at time $t$ arrive with a
\emph{distribution} of phase angles $\phi$, each carrying a
(normalized, transverse) magnetic moment $e^{i \phi}$. The transverse
magnetization then results, by construction, from \flexmath{align}{\MM
  = {\textstyle \int }\dd \phi \, \PXP e^{i\phi}}{eq:mag.phase}{,}
where $\PXP$ denotes the probability density for the joint event of a
random-walker having accumulated a phase of $\phi$ and ending up at
location $\vec{x}$ at time $t$. As a fundamental quantity, therefore,
the magnetization density $\MM$ lacks access to the randomness of the
phase variable $\phi$.

We henceforth refer to $\PXP$ as the \emph{\PPDxlong} (\PPD), and
propose its time evolution equation as a more complete alternative to
the Bloch-Torrey equation for the transverse magnetization density. We
start by describing how the evolution equation emerges, followed by
its analytical solution to tractable cases of relevance.

\subsection*{Evolution as a Fokker-Planck equation}

The time evolution of $\PXP$ hinges on the inclusion of the phase
$\phi$ along with position $\vec{x}$ in the list of random variables
pertaining to the problem, achieved as follows. Denoting by
\flexmath{align} {\OMGx = -\gamma \BZ}{eq:omega.b}{,} the field of
precession rate (in excess of $\omega_0$) imposed on the spins by the
manipulation of magnetic fields, a random-walker accumulates the angle
\flexmath{align}{\phi (t) = \int^t_0 \dd \tau \, \omega \vec{(}
  \vec{x}(\tau), \tau \vec{)}}{eq:phase.omega}{,} by time $t$ along
the trajectory $\vec{x}(\cdot)$. Following the procedure
\cite{RiskenBook} of connecting stochastic trajectory (Langevin)
equations to ensemble evolution (Fokker-Planck) equations, the
proposed evolution equation of $\PXP$ is obtained through an
augmentation of the (Smoluchowski) equation for Brownian motion by an
advective term along the $\phi$ coordinate as
\begin{align}
  \del_t \PXP = \Difop \PXP - \OMGx \del_\phi \PXP \,. \label{eq:PPD.evol}
\end{align}
\Eqref{eq:PPD.evol} has the form of a classical counterpart to the
density matrix evolution in Ref.\ \cite{Cates88}. We note that if
$\phi$ is defined on the entire real line, the periodized function
\flexmath{align} {\PXPper = \sum_{n=-\infty}^{\infty} \pxp (\vec{x},
  \phi + 2\pi n | t)\,, \quad (n \in \mathbb{Z})} {eq:periodized}{}
obeys \Eqref{eq:PPD.evol} with $2\pi$-periodicity along $\phi$. Either
function/definition may be adopted for convenience.

While we solve \Eqref{eq:PPD.evol} for specific cases later, the
action of the individual terms of the equation can be described
qualitatively here. As time goes on, the advective operator $-\OMGx
\del_\phi$ streams the probability in the neighborhood of $\vec{x}$
along the $\phi$ direction, while the operator $\Difop$ strives for
${e}^{\UU(\vec{x})} \PXP$ to lose $\vec{x}$-dependence. The latter
implies that as time goes on, the solution tends toward a family of
functions, \flexmath{align} {\PXP = {e}^{-\UU(\vec{x})} {f} (\phi|t)}
{eq:st.fam}{.} As $\Difop {e}^{-\UU(\vec{x})}=0$, this form yields
\flexmath{align} {\del_t {f} (\phi|t) = - \OMGx \del_\phi {f}
  (\phi|t)} {eq:PPD.evol.st}{,} via \Eqref{eq:PPD.evol}. Here, two
situations need to be distinguished: the precession field $\OMGx$
being spatially uniform or not. In the nonuniform case, there is no
way to balance the $\vec{x}$ dependence in the equation other than all
terms vanishing, implying ${f}(\phi|t)$ is a constant. Hence, when
$\OMGx$ is spatially non-uniform, every initial density tends toward
the unique $\phi$-independent stationary solution $\ssdxp
(\vec{x},\phi) \sim {e}^{-\UU(\vec{x})}$. With spatially uniform
$\OMGx=\omega(t)$, on the other hand, one finds ${f}(\phi|t) = {f}
\vec{(} \phi - \int_{t_0}^t \dd \tau \, \omega(\tau) | t_0 \vec{)}$,
where $t_0$ is an inconsequential time large enough that $\Difop \pxp
(\vec{x},\phi|t_0) \approx 0$. This describes a long-time density
$\PXP = {e}^{-\UU(\vec{x})} {f} (\phi|t)$ that slides ``rigidly''
along the $\phi$ direction. In this case with spatially uniform
precession field, therefore, only \emph{some} initial densities reach
the stationary state $\ssdxp (\vec{x},\phi)\sim {e}^{-\UU(\vec{x})}$;
those that reach it before ${e}^{\UU(\vec{x})} \PXP$ becomes
$\vec{x}$-independent.

Finally, note that although the physical problem may demand other
boundary conditions, the validity of the stationarity arguments above
require vanishing probability current
\begin{align}
  \vec{J} = - e^{- \UU(\vec{x})} \DD (\vec{x}) \cdot
    \nabla \frac {\PXP}{e^{- \UU(\vec{x})}} + \OMGx \PXP \,, \label{eq:current}
\end{align}
through the boundary of the
space spanned by $(\vec{x},\phi)$, which may be at infinity.

\subsection*{Connection to Bloch-Torrey equation}

There is an illuminating connection between the transverse BTE and the
\PPD\ evolution. This is seen by Fourier transforming
over $\phi$ in \Eqref{eq:PPD.evol}:
\begin{align}
  \del_t \PHXP = \Difop \PHXP + {i} \lambda \OMGx \PHXP \,, \label{eq:PPD.evol.F}
\end{align}
where $\lambda$ is the conjugate of $\phi$ via the relation
\flexmath{align}{\PHXP = \textstyle\int \dd \phi \, \PXP
  e^{i\lambda\phi}}{eq:PPD.F}{.} \Eqref{eq:PPD.evol.F} is the BTE
\eqref{eq:BTE} under a rescaled magnetic field $\BZ \to \lambda
\BZ$. Hence, \Eqref{eq:PPD.evol} is equivalent to a \emph{family} of
BTEs \eqref{eq:PPD.evol.F} spanned by the overall strength $\lambda$
of the magnetic field. This would imply that the magnetization density
is one single Fourier mode of the \PPD\ at $\lambda=1$. That is, $\MM
= \phxp (\vec{x}, 1| t)$, which is of course identical to its
definition \eqref{eq:mag.phase}.

Note also that the \PPD\ could be expressed as
\flexmath{align} {\PXP = \Big< \delta \big( \vec{x}-\vec{x}(t) \big)
   \delta \big( \phi - {\textstyle \int^t_0 \dd \tau \omega
    \vec{(} \vec{x}(\tau),\tau \vec{)} } \big) \Big>} {eq:FeynKac} {,}
which is true by construction, where the average is over the ensemble
of stochastic trajectories $\vec{x}(\cdot)$. Then by the theorem of
Feynman-Kac \cite{Feynman48,Kac49}, \Eqref{eq:PPD.evol.F} can be
obtained.

\subsection*{Free diffusion under field gradient}

Free homogeneous diffusion where the longitudinal magnetic field
(precession field) varies linearly, $\OMGx = -\gamma \gw(t) \cdot
\vec{x}$, is a case which permits solution of \Eqref{eq:PPD.evol}
analytically, for instance, via the transformation \flexmath{inline}
{\PXP = e^{\vec{x} \cdot \qq (t) \del_\phi} \psi (\vec{x}, \phi |
  t)}{}{,} where the wave vector \flexmath{align} {\qq (t) = \gamma
  {\int^t_0} \dd \tau \, \gw (\tau)}{eq:q.free}{.} For the
spatially-uniform and angularly-coherent initial condition $\pxp
(\vec{x},\phi|0) \propto \delta (\phi)$, one finds the Gaussian \PPD\
\begin{align}
  \PXP = \ssdx \frac {\exp \!\left[
      {\frac {-\left[\phi + \vec{x} \cdot \qq (t) \right]^2}
      {4 \int^t_0 \dd \tau \, \qq(\tau)^\mT \DD \qq(\tau)}} \right] }
  {\sqrt{4 \pi \int^t_0 \dd \tau \, \qq(\tau)^\mT \DD \qq(\tau)}} \label{eq:PPD.free} \,,
\end{align}
where $\ssdx$ is the uniform stationary density in free
space.\footnote{Of course, the proportionality constant $\ssdx$ is a
  vanishing quantity. Despite this, we treat it as a legitimate
  probability density, in the same sense plane waves are dealt with in
  quantum mechanics.} The matrix $\BB (t) = \int^t_0 \dd \tau \, \qq
(\tau) \qq (\tau)^\mT$ is commonly dubbed the ``diffusion-weighting
tensor'' in diffusion-weighted NMR acquisitions that probe the
anisotropy of the movement of water molecules in tissues and
materials. One observes that the variance $ 2 \int^t_0 \dd \tau \,
\qq(\tau)^\mT \DD \qq(\tau) = 2 \Tr \! \left[ \BB(t) \DD \right]$ is a
non-decreasing function of time, remaining constant only on intervals
where $\qq(t)=0$. Hence the distribution $\PXP$ keeps approaching
uniformity along the $\phi$ coordinate for as long as the past average
of the gradient $\gw (t)$ is nonzero, even if $\gw (t)$ itself is
``turned off.'' This is a consequence of the infinity of the $\vec{x}$
domain.

It is verified easily that the magnetization density
\eqref{eq:mag.phase} \flexmath{align} {\MM = \ssdx e^{-i \vec{x} \cdot
    \qq (t)} e^{-\Tr \left[ \BB(t) \DD \right]}} {eq:mag.free}{,} and
the signal amplitude \eqref{eq:signal} \flexmath{align} {\EE (t) =
  e^{-\Tr \left[ \BB(t) \DD \right]}} {eq:signal.free}{,} provided
$\qq (t) =0$ (and $\EE=0$ otherwise), as are well-known
\cite{StejskalTanner65,Karlicek80}. We see that the spread $2 \Tr \!
\left[ \BB(t) \DD \right]$ of the distribution of phase angles is what
determines the attenuation of the signal.

\subsection*{Harmonically confined diffusion under field gradient}

Whereas the diffusion process gets quite intractable analytically
when it takes place inside finite domains or in the presence of
arbitrary tissue inhomogeneities, approximating the confinement by a
harmonic (Hookean) force eases the problem while retaining the
anisotropy of motion, as well as its finite extent \cite{Yolcu16}.

The \PPD\ for the corresponding (dimensionless)
potential $\UU (\vec{x}) = (1/2) (\vec{x}-\vec{x}_0)^\mT \CC
(\vec{x}-\vec{x}_0)$, where $\CC$ is an anisotropic tensor of spring
constants (scaled by $1/\kB T$), is most easily obtained by invoking
\Eqref{eq:FeynKac}. Thanks to the process having a Gaussian
probability measure, this path integral can be evaluated without much
challenge. We obtain
\begin{subequations} \label{eq:PPD.conf}
\begin{align}
  \PXP = \ssdx^\CC (\vec {x}) \frac
       {\exp \! \left[{ \frac {-\left[ \phi + \vec{x}_0 \cdot \qq (t)
                 + (\vec{x}-\vec{x}_0) \cdot \QQ (t) \right]^2}
      {4 \int^t_0 \dd \tau \, \QQ (\tau)^\mT \DD \QQ (\tau)} } \right] }
       {\sqrt{4 \pi \int^t_0 \dd \tau \, \QQ (\tau)^\mT \DD \QQ (\tau)}}
       \,, \label{eq:PPD.conf.sub}
\end{align}
where \flexmath{align} {\QQ(t) = \gamma {\textstyle \int^t_0} \dd \tau
  \, {e}^{-\DD\CC|t-\tau|} \gw (\tau)} {eq:q.conf}{,} and
\flexmath{align} {\ssdx^\CC (\vec{x}) = \frac {\exp \!\left[ {-\frac12
    (\vec{x}-\vec{x}_0)^\mT \CC
    (\vec{x}-\vec{x}_0)} \right]}{\sqrt{\det(2\pi\CC^{-1})}}}{eq:ssd.conf}{}
\end{subequations}
is the stationary ($t \to \infty$) distribution of the process
$\vec{x}(\cdot)$, due to the assumed initial condition $\pxp (\vec{x},
\phi |0) = \ssdx^\CC (\vec{x}) \delta(\phi)$. Moreover, the tensors
$\DD$ and $\CC$ of diffusivity and confinement are assumed to share
the same set of principal directions. It is verified easily that the
confined solution \eqref{eq:PPD.conf} tends to the force-free
solution \eqref{eq:PPD.free} as the spring constants vanish ($\CC \to
0$).

The variance $2 \int^t_0 \dd \tau \, \QQ (\tau)^\mT \DD \QQ (\tau)$ is
manifestly non-negative and a non-decreasing function of $t$. Thus the
distribution keeps spreading along the $\phi$ coordinate as long as
the exponentially-weighted past average \eqref{eq:q.conf} of the
gradient $\gw (t)$ is nonzero. The spread does not go on forever like
the force-free case though. It effectively stops several multiples of
the largest eigen-time of $(\DD \CC)^{-1}$ after the gradient $\gw
(t)$ is turned off.

The magnetization density follows as \flexmath{align} {\MM = \ssdx^\CC
  (\vec{x}) {e}^{- \int^t_0 \dd \tau \, \QQ(\tau)^\mT \DD \QQ (\tau)}
  {e}^{-{i} \QQ(t) \cdot (\vec{x}-\vec{x}_0)-{i} \qq(t) \cdot
    \vec{x}_0} }{eq:mag.conf}{.} Namely, a Gaussian wave packet of
covariance $\CC$ whose wave vector is given by \Eqref{eq:q.conf} and
attenuation determined by the spread $2 \int^t_0 \dd \tau \,
\QQ(\tau)^\mT \DD \QQ (\tau)$ of $\PXP$ along $\phi$.

\begin{figure}
  \includegraphics[width=.99\columnwidth]{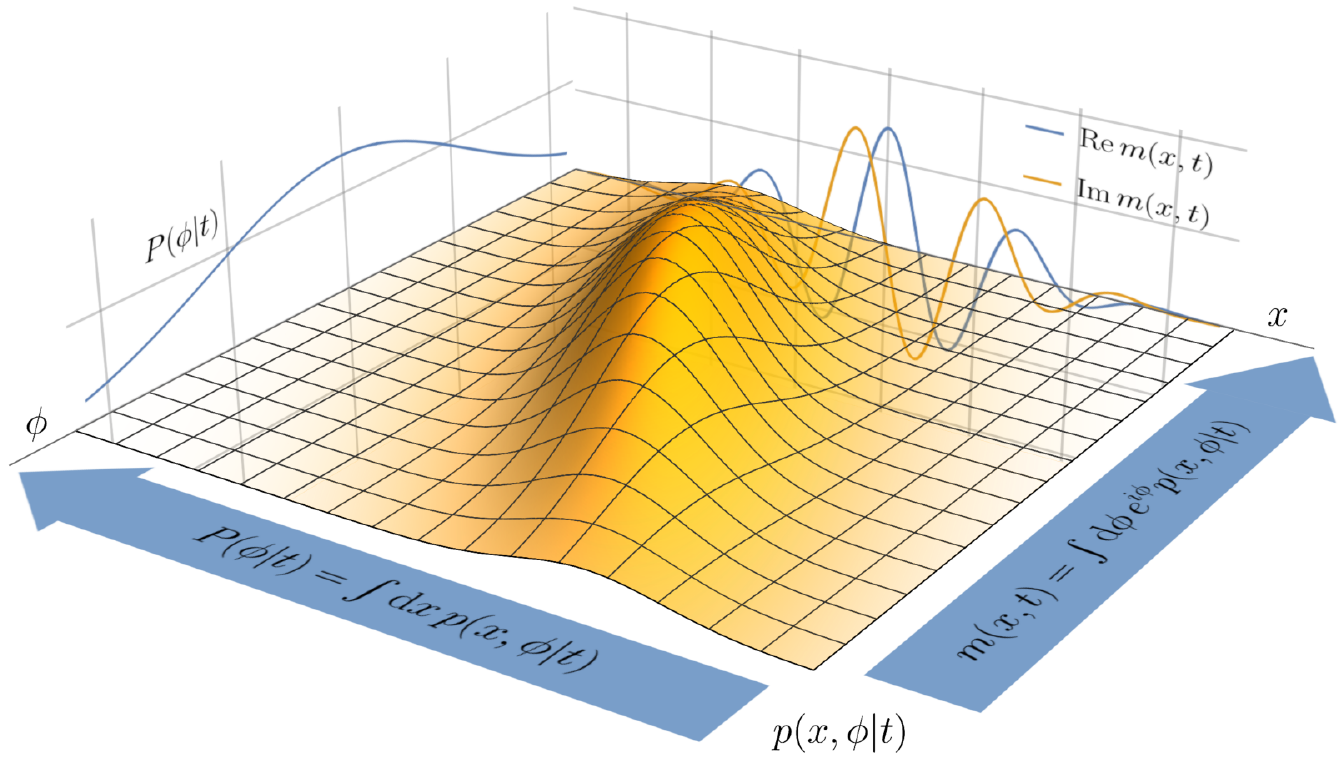}
  \caption{The solution \eqref{eq:PPD.conf} for representative
    parameter values. The \PPD\ $\PXP$ is the fundamental quantity out
    of which the phase density $\PP(\phi|t)$ and magnetization density
    $\MM$ emerge upon marginalization/averaging. \label{fig:marginal}}
\end{figure}

\subsection*{Discussion}

The NMR signal can be expressed as $\EE (t) = \int \dd \phi \,
{e}^{{i}\phi} \PP (\phi|t)$, and in turn $\PP (\phi|t) = \int \dd^3x
\, \PXP$; see Fig.\ \ref{fig:marginal}. Hence the evolution
\eqref{eq:PPD.evol} of the \PPD\ $\PXP$ furnishes
insight into how the (global) phase distribution $\PP (\phi|t)$
develops.

The phase distribution $\PP (\phi|t)$ is in general not Gaussian,
although it is often approximated to be so, be it directly
\cite{Neuman74,Murday68,Grebenkov07}, or indirectly via asserting that
it obeys diffusion along $\phi$ \cite{Lin15}. Recently, it was
proposed that approximating the conditional probability $\pxp (\phi |
\vec{x}, t)$ to be Gaussian rather than the global phase distribution
$\PP (\phi|t)$ is closer to reality \cite{Ziener18}. The nonzero
cumulants, of which there are two, were furthermore stated to obey
certain partial differential equations. These equations, as well as
those of higher nonvanishing cumulants, are encompassed by the
\PPD\ evolution \eqref{eq:PPD.evol} upon substitution of $\PXP = \pxp
(\phi| \vec{x},t) / \px (\vec{x}|t)$:
\begin{align}
  \del_t \pxp (\phi |\vec{x}, t) &= \Difop_* \pxp (\phi|
  \vec{x},t) - \OMGx \del_\phi \pxp (\phi |\vec{x},t) \,
\end{align}
with the operator
\begin{align}
  \Difop_* &= \ssdx^{-1} (\vec{x}) \Difop \ssdx (\vec{x}) + 2 \left(
  \nabla \ln \frac {\px (\vec{x}|t)}{ \ssdx (\vec{x})} \right) \cdot \nabla \,,
\end{align}
and $\ssdx (\vec{x}) \propto {e}^{-\UU(\vec{x})}$, generalizing them
to the non-stationary non-Gaussian
case.\footnote{Defining the $n$'th conditional moment
  $\mu_n (\vec{x}, t) = \int \dd \phi \, \frac {\PXP} {\ssdx
    (\vec{x})} \phi^n$ alternatively yields the coupled system of
  equations $$ \del_t \mu_n (\vec{x},t) = \Difop_* \mu_n (\vec{x},t) +
  n \OMGx \mu_{n-1} (\vec{x},t) \,,$$ the first two of which were
  given in Ref.\ \cite{Ziener18} for the stationary case with
  vanishing potential.} Note that a Hookean confinement
\eqref{eq:PPD.conf}, including the limit $\CC \to 0$, constitutes the
general problem (under a linear field gradient, that is) where the
distribution $\pxp (\phi |\vec{x}, t)$ of phase angles is
\emph{actually} Gaussian, since it uses the most general stationary
Gaussian process in \Eqref{eq:FeynKac}.

Upon the simple linear coordinate transformation $\tilde\phi = \lambda
\phi$, the probability density $\tilde\pxp (\vec{x}, \tilde\phi |t) =
\lambda^{-1} \PXP$ is found to obey \Eqref{eq:PPD.evol} with the
replacement $\OMGx \to \lambda \OMGx$.
In other words, once the \PPD\ $\PXP$ evolving under
field $\OMGx$ is known, solutions under $\lambda \OMGx$ for all
$\lambda$ follow trivially as $\lambda^{-1} \pxp (\vec{x},
\lambda^{-1} \phi |t)$. In contrast, magnetization densities $\MM$
solving the transverse Bloch-Torrey equation (BTE) do not exhibit such
a feature in any obvious way \cite{Stoller91,Herberthson17JCP}. This
is closely related to the previously explained fact that $\PXP$
comprises the solutions to an entire family of transverse BTEs.

Finally, we note that even though the precession field $\OMGx$ was
written as a function of position, a dependence on the phase angle
$\phi$ can be accounted for simply by swapping the last term in
\Eqref{eq:PPD.evol} with $-\del_\phi \OMG \PXP$. While it is not the
typical scenario, this version may be applicable to experiments where
quasi-instantaneous RF pulses rotate spins around an axis within the
transverse plane until they are back in the transverse plane at an
angle that is a function of its initial value.

In closing, we introduced the \PPD\ and its evolution
\eqref{eq:PPD.evol} as a framework for studying the dynamics of
magnetic moments transverse to a polarizing field with significant
conceptual benefits over the traditional BTE.

\begin{acknowledgments}
  The authors thank Hans Knutsson for stimulating discussions, and
  acknowledge the Swedish Foundation for Strategic Research AM13-0090,
  the Swedish Research Council 2016-04482, Linköping University Center
  for Industrial Information Technology (CENIIT), VINNOVA/ITEA3 17021
  IMPACT, and National Institutes of Health P41EB015902, R01MH074794.
\end{acknowledgments}

\end{document}